\documentclass[twocolumn,english,aps,prl,groupedaddress,superscriptaddress]{revtex4}
\usepackage{graphicx,epsfig,units}
\usepackage{siunitx}
\usepackage{xcolor} 
\usepackage{soul} 
\usepackage{amsmath,amsfonts,mathrsfs,amsbsy,bm,babel}

\usepackage{dcolumn}
\usepackage{mciteplus}
\usepackage{euscript}
\usepackage{multirow}
\usepackage{textcomp}
\usepackage{gensymb}
\usepackage{dblfloatfix}
\usepackage{float}
\usepackage{enumitem}
\usepackage{sepfootnotes}
\usepackage{mhchem}

\begin{document}
\title{Defect-driven ferrimagnetism and hidden magnetization in MnBi$_2$Te$_4$}

\author{You Lai}
\email{youlai0514@gmail.com}
\affiliation{Los Alamos National Laboratory, Los Alamos, NM, 87545, USA}

\author{Liqin Ke}
\affiliation{Ames Laboratory, Ames, IA, 50011, USA}

\author{Jiaqiang Yan}
\affiliation{Oak Ridge National Laboratory, Oak Ridge, TN, 37831, USA}

\author{Ross D. McDonald}
\affiliation{Los Alamos National Laboratory, Los Alamos, NM, 87545, USA}

\author{Robert J. McQueeney}
\email{rmcqueen@ameslab.gov}
\affiliation{Ames Laboratory, Ames, IA, 50011, USA}
\affiliation{Department of Physics and Astronomy, Iowa State University, Ames, IA, 50011, USA}

\date{\today}

\begin{abstract}
MnBi$_2$Te$_4$ (MBT) materials are promising antiferromagnetic topological insulators where field driven ferromagnetism is predicted to cause a transition between axion insulator and Weyl semimetallic states. However, the presence of antiferromagnetic coupling between Mn/Bi antisite defects and the main Mn layer can reduce the low-field magnetization, and it has been shown that such defects are more prevalent in the structurally identical magnetic insulator MnSb$_2$Te$_4$ (MST). We use high-field magnetization measurements to show that the magnetization of MBT and MST occur in stages and full saturation requires fields of~$\sim$~60 Tesla. As a consequence, the low-field magnetization plateau state in MBT, where many determinations of quantum anomalous Hall state are studied, actually consists of ferrimagnetic septuple blocks containing both a uniform and staggered magnetization component.  
\end{abstract}
\maketitle
\section{Introduction}

Shortly after MnBi$_2$Te$_4$ (MBT) was recognized to be the first antiferromagnetic topological insulator~\cite{Otrokov19}, it was realized that it may be susceptible to various mechanisms of chemical disorder, such as antisite mixing and vacancies that could possibly affect the magnetic structure. Ideally, MBT compounds consist of blocks of stacked triangular layers in the sequence Te-Bi-Te-Mn-Te-Bi-Te (a septuple block).  The Mn layer in each septuple block is ferromagnetic (FM) and alternating block-to-block magnetization forms an A-type antiferromagnetic (AF) structure. However, in the structurally identical magnetic insulator MnSb$_{2}$Te$_{4}$ (MST), antisite and vacancy defects are prevalent (at the level of 10 -- 20 \%) and their effects are easily observable. Neutron~\cite{Murakami19, Liu20} and x-ray~\cite{zhou20} diffraction measurements in MST definitively show that significant antisite mixing occurs between Mn and Sb sites. Neutron data also confirms that antisite magnetic Mn ions in the Sb layer couple antiferromagnetically to the main Mn layer, forming ferrimagnetic septuple blocks.~\cite{Murakami19, Liu20}  In MST, it has been shown that ferrimagnetic blocks can stack either AF or FM depending on the synthesis conditions that primarily affect the defect concentrations.~\cite{Liu20}

The presence of magnetic defects in the magnetic topological insulator MBT and its effect on its magnetic, topological, and surface state properties is not fully understood. Mn/X antisite defects in both MBT and MST can be enumerated by this site-specific chemical formula (Mn$_{1-2x-y}$X$_{2x+y}$)(X$_{1-x}$Mn$_{x}$)$_{2}$Te$_4$ where ${\rm X}=$ Bi or Sb, $x$ is the antisite mixing concentration (which preserves the overall stoichiometry), and $y$ is the concentration of additional X ions on the Mn site. In MBT, various experimental methods such as elemental analysis,~\cite{Fan20} x-ray diffraction~\cite{Zeugner19}, neutron diffraction~\cite{Yan19, Ding20}, and scanning tunneling microscopy~\cite{Yuan20, Liang20} provide some insight into the values of $x$ and $y$ in real materials. A survey of the literature, as shown in Table~\ref{table1}, finds a large range of values for $x = 0 - 0.08$ and $y = 0 - 0.15$ which are dependent on various assumptions, such as vacancy concentration and stoichiometry.  However, these data may also suggest that the chemical configuration is highly dependent on the synthesis conditions.  

Experimental evidence of antisite Mn defects is also found in the reported values of the saturation magnetization conducted in applied fields of less than 12~T. In MBT, the saturation magnetization at 10~T is approximately 4~$\mu_{\rm B}$ per Mn which is less than the expected value of 4.5 -- 5~$\mu_{\rm B}$ obtained from neutron scattering measurements.~\cite{Yan19} By comparison, the saturation magnetization in MST is around 2~$\mu_{\rm B}$ \cite{Yan19_MST} due to strong ferrimagnetic compensation. For both materials, magnetization measurements in much higher applied fields should uncover the missing magnetization. 

Here we use high-field magnetization to study the response of magnetic ions and their defects in MBT and MST up to 60$-$70~T, where, in both cases, the missing magnetization is recovered.  In MBT, the high-field magnetization data reveal two plateaus.  As mentioned above, the first plateau occurs near 10~T and has a magnetization slightly less than $ 4~\mu_{\rm B}$. The second plateau saturates around 50~T, recovering the full magnetic moment of approximately $4.6~\mu_{\rm B}$ per Mn. Classical Monte Carlo~(CMC) simulations assuming random antisite Mn defects at a concentration of $x = 0.038$ with AF coupling to the main layer are consistent with experimental data for MBT. This combination of data and simulations confirm the ferrimagnetism model also for MBT, where a single septuple block remains ferrimagnetic at fields less than 10~T. For MST, both FM and AF samples were studied and magnetization data find similar levels of antisite mixing ($x$). However, the AF samples have a much higher concentration of magnetic vacancies ($y$). The magnetization data and simulations also confirm strong AF coupling of antisite Mn ions to the main Mn layer in MBT and both FM and AF MST. We compare these experimental results to predictions of the intrablock magnetic couplings using density-functional theory (DFT). Surprisingly, the DFT results predict \textit{ferromagnetic} coupling of isolated Mn defects in the Bi or Sb layers to the main Mn layer. However, the introduction of correlated antisite defects is found to promote the observed AF interblock coupling and ferrimagnetism.

\begin{table}
\centering
    \begin{tabular}{c|c|c|c}
          MXT & $x$ & $y$  & method  \\
         \hline\hline
    MBT 		& 0.01(1)         		& 0.18(1)  	& neutron: \cite{Ding20} \\
    MBT 		& 0.057 (1)    	& 0.15 	& x-ray (Model 1): \cite{Zeugner19} \\
    MBT 		& 0.088 (1)    	& 0    	& x-ray (Model 2): \cite{Zeugner19} \\
    MBT 		& 0.05		& -     	& STM: \cite{Yuan20} \\
    MBT 		& 0.03(0.2)     	& 0     	& STM: \cite{Huang20} \\
    MBT 		& 0.038      	& 0        	& This work \\
    \hline\hline
    MST-FM    	& 0.17      		& 0     	& neutron: \cite{Murakami19} \\
    MST-AF    	& 0.129(2) 	& 0.154   	& neutron/EDX: \cite{Liu20} \\
    MST-FM(1)   & 0.150(2) 	& 0.065   	& neutron/EDX: \cite{Liu20} \\
    MST-FM(2)   & 0.158(3) 	& 0.01   	& neutron/EDX:  \cite{Liu20} \\
    MST-AF	& 0.13        	& 0.22      	& This work \\
    MST-FM	& 0.17          	& 0      	& This work \\
    \hline\hline
    \end{tabular}
    \caption{ Chemical and magnetic configurations of MBT and MST, as reported in different publications, according to the formula (Mn$_{1-2x-y}$X$_{2x+y}$)(X$_{1-x}$Mn$_{x}$)$_{2}$Te$_4$.}
    \label{table1}
\end{table}

\section{Experimental details}

MnBi$_2$Te$_4$ single crystals were grown by the flux method described in Ref.~\cite{Yan19} with a reported N\'eel temperature of $T_{\rm N}=$ 25~K . Elemental analysis was performed on freshly cleaved surfaces using a Hitachi TM-3000 tabletop electron microscope equipped with a Bruker Quantax 70 energy dispersive x-ray system (EDS). Our MBT sample was determined to have a stoichiometry of Mn$_{0.99(1)}$Bi$_{2.01(2)}$Te$_{4.00(2)}$. Scanning tunneling microscopy measurements \cite{Huang20} show there are $x \approx 0.03$ Mn on the Bi site (Mn$_{\rm Bi}$) and $ \sim $0.2\% Bi$_{\rm Te}$ antisite defects. Single crystals of both AF ($T_{\rm N}=$ 19~K) and FM ($T_{\rm C}=$ 28~K) variants of MnSb$_2$Te$_4$, MST-AF and MST-FM respectively, were grown using the method in Ref.~\cite{Liu20}. Elemental analysis finds the MST-FM sample to have a nearly stoichiometric composition of Mn$_{1.00(4)}$Sb$_{2.09(2)}$Te$_{3.91(2)}$ without Mn deficiency. For the MST-AF sample, elemental analysis finds heavy Mn deficiency with a composition of Mn$_{0.778(5)}$Sb$_{2.222(6)}$Te$_{4.00(1)}$.~\cite{Liu20}

High-field magnetization measurements were conducted at the Pulsed-Field Facility of the National High Magnetic Field Laboratory at Los Alamos National Laboratory using an extraction magnetometer in a 70~T short-pulse magnet. The extraction magnetometer is home-made and consists of a 1.5~mm bore, 1.5~mm long, 1500-turn compensated-coil susceptometer, constructed from 50 gauge high-purity copper wire. The bulk samples are mounted within a 1.3~mm diameter ampoule that can be moved in and out of the coil. When a sample is within the coil, the time~($t$) varying magnetization signal $V(dM/dt)$ is induced by the $dB/dt$ of the pulsed magnet.~\cite{Goddard08}

Numerical integration is used to evaluate the sample magnetization $M$. Accurate values of $M$ are obtained by subtracting empty coil data from that measured under identical conditions with the sample present.~\cite{Goddard08} The magnetization data are then normalized and calibrated from independent low-field magnetization data in measured fields up to 12~T that were collected using the AC option of a Quantum Design Physical Property Measurement System. The low field magnetization data up to 7~T for MST were collected using a Quantum Design Magnetic Property Measurement System. All magnetization data are reported in Bohr magnetons per formula unit ($\mu_{\rm B}/$fu) or per Mn ($\mu_{\rm B}/$Mn), as appropriate, using EDS determined stoichiometries.

\section{Analysis of magnetization data for MBT}

\begin{figure}
\includegraphics[width=3. in]{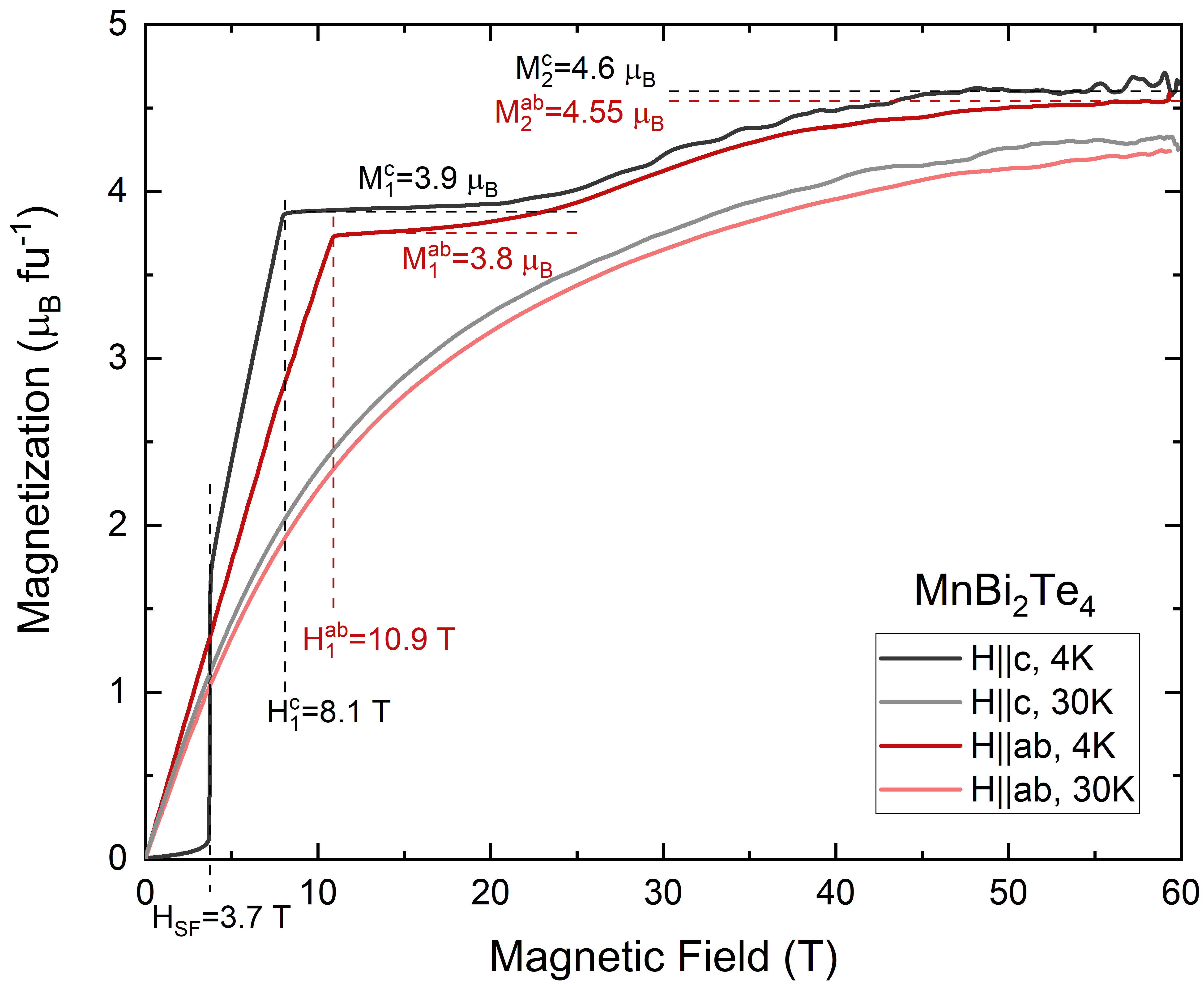}
\caption{High-field magnetization data of Mn$_{0.99(1)}$Bi$_{2.01(2)}$Te$_{4.00(2)}$ at 4~K and 30~K with the field applied either along the $c$-axis or in the $ab$-plane. }
\label{data}
\end{figure}

Fig.~\ref{data} gives an overview of the high-field magnetization data for MBT. For fields applied along $c$ and starting in the zero-field AF ordered phase at $T=$ 4~K, the magnetization jumps from the A-type AF phase into the canted spin-flop phase near $\mu_0 H_{\rm SF}=3.7$~T. In the spin-flop phase, the magnetization increases linearly up to a field of $\mu_0 H_1^c \approx$ 8~T, where the magnetization reaches a plateau with $M_1^c \approx 3.9~\mu_{\rm B}/$fu. Above $\mu_0 H_1^c$, the magnetization smoothly increases to a second magnetization plateau with $M_2^{c}$ = 4.6~$\mu_{\rm B}/$fu near 50~T. For the field applied in the $ab$ plane, the low-field magnetization evolves linearly up to a field of $\mu_0 H_1^{ab}=$ 10.9~T due to coherent moment rotation from the $c$-axis into the $ab$ plane. Above $\mu_0 H_1^{ab}$, the evolution of the magnetization is similar to that for $H \parallel c$, reaching magnetization plateau of $M_2^{ab}=$ 4.55~$\mu_{\rm B}/$fu near 60~T. We propose that the evolution to second plateau is due to the gradual spin flip of antisite Mn ions residing on the Bi site (Mn$_{\rm Bi}$) ions.

For both directions, we assume that the magnetization of 4.55 -- 4.6~$\mu_{\rm B}/$fu at 60~T represents full magnetic saturation of the MBT sample.  EDS results find our MBT sample to be stoichiometric within error, so this value represents the local Mn moment size, denoted by $m_0$. The value of $m_0$ is consistent with neutron diffraction~\cite{Yan19, Ding20} and DFT calculations~\cite{Otrokov19, Yan19_MST} that find $m_0$ to be slightly reduced from the $S=5/2$ local moment value of 5~$\mu_B$ due to orbital hybridization. For subsequent analysis of MBT, we assume a Lande $g$-factor of 1.84 instead of 2 to account for the reduced moment.

Figure~\ref{data} also shows the magnetization of MBT for temperatures above $T_{\rm N}$. Here, the magnetization evolves smoothly for either field direction without reaching saturation at 60~T. Slight differences in the both the 4~K and 30~K magnetization for the two field directions suggest a weakly anisotropic Lande $g$-factor.

To understand the magnetization data, we develop a simple Heisenberg model for MBT and employ CMC simulations. We show below that CMC simulations are a reasonable approximation at the lowest temperatures where spin fluctuations are small. The Heisenberg Hamiltonian has the following parameters; $J$ is the FM exchange coupling within the Mn layer, $J_c$ is the AF coupling between septuple blocks, $J'$ is the AF coupling between antisite Mn$_{\rm Bi}$ and the main Mn$_{\rm Mn}$ sites within a septuple block (intrablock coupling), and $D$ is the single-ion (uniaxial) anisotropy.  The $J_c$ and $D$ parameters can be estimated from the low-field saturation anisotropy using the Stoner-Wohlfarth model~\cite{swmodel}
\begin{equation}
     SJ_c = \frac{1}{4z}g \mu_{\rm B} (H_1^{ab}+H_1^{c})
     \label{eqnJc}
 \end{equation}
  \begin{equation}
     SD = \frac{1}{4}g \mu_{\rm B} (H_1^{ab}-H_1^{c})
     \label{eqnD}
 \end{equation}
where $S=5/2$, and $z=6$ is the number of Mn nearest-neighbors in adjacent septuple blocks.  

From these relations, we obtain values of $SD=$ 0.075~meV and $SJ_c =$ 0.085~meV (see Table \ref{table2}). These parameters also predict a critical spin-flop field of $g\mu_{\rm B}H_{\rm SF} = 2SD\sqrt{zJ_c/D-1} =~$3.4~T close to experimental results. Subsequent simulations (not shown) indicate that the low-field transitions depend only weakly on $J'$ and the defect concentration.This gives confidence that our estimates of $D$ and $J_c$ are reasonable and these parameters are fixed for subsequent CMC simulations. The chosen value of the intralayer exchange $SJ$ = 0.35~meV is consistent with the N\'eel temperature and also INS data \cite{Li20}, although this parameter is not critically important for modeling the low temperature magnetization.

\begin{table}
\centering
    \begin{tabular}{c|c|c|c}
          ~~ 			& ~MBT~ & ~MST (AF)~ & ~MST (FM)~  	\\
         \hline\hline
    $H_1^c$ (T)		& 8.1		& 0.5     		& $\sim$0	\\
    $H_1^{ab}$ (T)	& 10.9 	& 2.2  		& 1.2 	\\
    \hline
    $SJ_c$ (meV)	& 0.085	& 0.01  		& $\sim$0	\\
    $SD$ (meV)		& 0.075	& 0.07		& 0.07 	\\
    $SJ'$ (meV)		& 1.2		& 2.1			& 2.1		\\
    \hline\hline
    \end{tabular}
    \caption{Estimates of the Heisenberg parameters of MBT and MST samples based on critical fields determined from magnetization data (Eqns.~\ref{eqnJc} and \ref{eqnD} for MBT and \ref{eqnJc_MST} and \ref{eqnD_MST} for MST).  $SJ'$ is determined from CMC simulations.}
    \label{table2}
\end{table}

CMC analysis of the magnetization data allows us to estimate the strength of the intrablock AF coupling ($J'$) between antisite Mn$_{\rm Bi}$ and Mn$_{\rm Mn}$ ions, but first we need to consider the potential chemical configuration of the Mn ions.  After assuming a Mn moment size, $m_{0}$, both $x$ and $y$ can be estimated from the low-field ~(M$_1$) and high-field~(M$_2$) magnetization plateaus respectively. 
 \begin{equation}
     x=\frac{M_2-M_1}{4m_0}
     \label{eqnx}
 \end{equation}
 \begin{equation}
     y = \frac{m_0-M_2}{m_0}
 \label{eqny}
 \end{equation}
The EDS results for our MBT samples indicate a nearly stoichiometric compound which appears fully saturated by 60~T. This allows us to simplify our MBT model by setting $y = 0$ and $M_2 = m_0$. The antisite concentration $x$ is easily estimated by the magnetization jump between low and high-field plateaus, $x \approx 0.04$ (see Fig.~\ref{data} and Table~\ref{table3}), which is consistent with the STM result.~\cite{Huang20}

\begin{table}
\centering
    \begin{tabular}{c|c|c|c}
           & ~MBT~ & ~MST (AF)~ & ~MST (FM)~  \\
         \hline\hline
    $M_1^c (\mu_{\rm B}$/fu) 	& 3.9		& 1.6     	& 1.8		\\
    $M_2^c (\mu_{\rm B}$/fu)	& 4.6		& 3.8  	& 4.9 	\\
    $m_0 (\mu_B)$			& 4.6		& 4.9		& 5.3	\\
    \hline
    $x$ (mag)				& 0.04	& 0.11	& 0.15	\\
    $x$ (MC)				&0.038	& 0.13	& 0.17	\\
    $y$ (mag)				& $\sim$0	& 0.24	& $\sim$0	\\
    $y$ (EDS)				& 0.01	& 0.22	& 0	 	\\
    \hline\hline
    \end{tabular}
    \caption{Estimates of the chemical configuration of (Mn$_{1-2x-y}$X$_{2x+y}$)(X$_{1-x}$Mn$_{x}$)$_{2}$Te$_4$ (MXT) samples based on magnetization (Eqns.~\ref{eqnx} and \ref{eqny}) and EDS data. Slight refinements of the antisite mixing concentration $x$ based on comparison to CMC simulations are also shown.}
    \label{table3}
\end{table}

Using these initial estimates, CMC simulations were performed using the UppASD atomic spin dynamics package \cite{UppASD}. The model consists of simulations on a lattice of $11\times11\times20$ crystallographic unit cells (7260 spins) with $T=$ 0.1~K. The low temperature is chosen to suppress spin fluctuations which have a different temperature evolution in the classical simulations as compared to the quantum nature of $S = 5/2$ spins. The crystallographic unit cell consists of three septuple blocks (three Mn layers and six Bi layers) with a probability of $x$ that a Bi site is randomly occupied by Mn (a Mn$_{\rm Bi}$ magnetic impurity) and $2x$ that a Mn site is occupied by Bi (a Bi$_{\rm Mn}$ magnetic vacancy).  

\begin{figure}
\includegraphics[width=3.3 in]{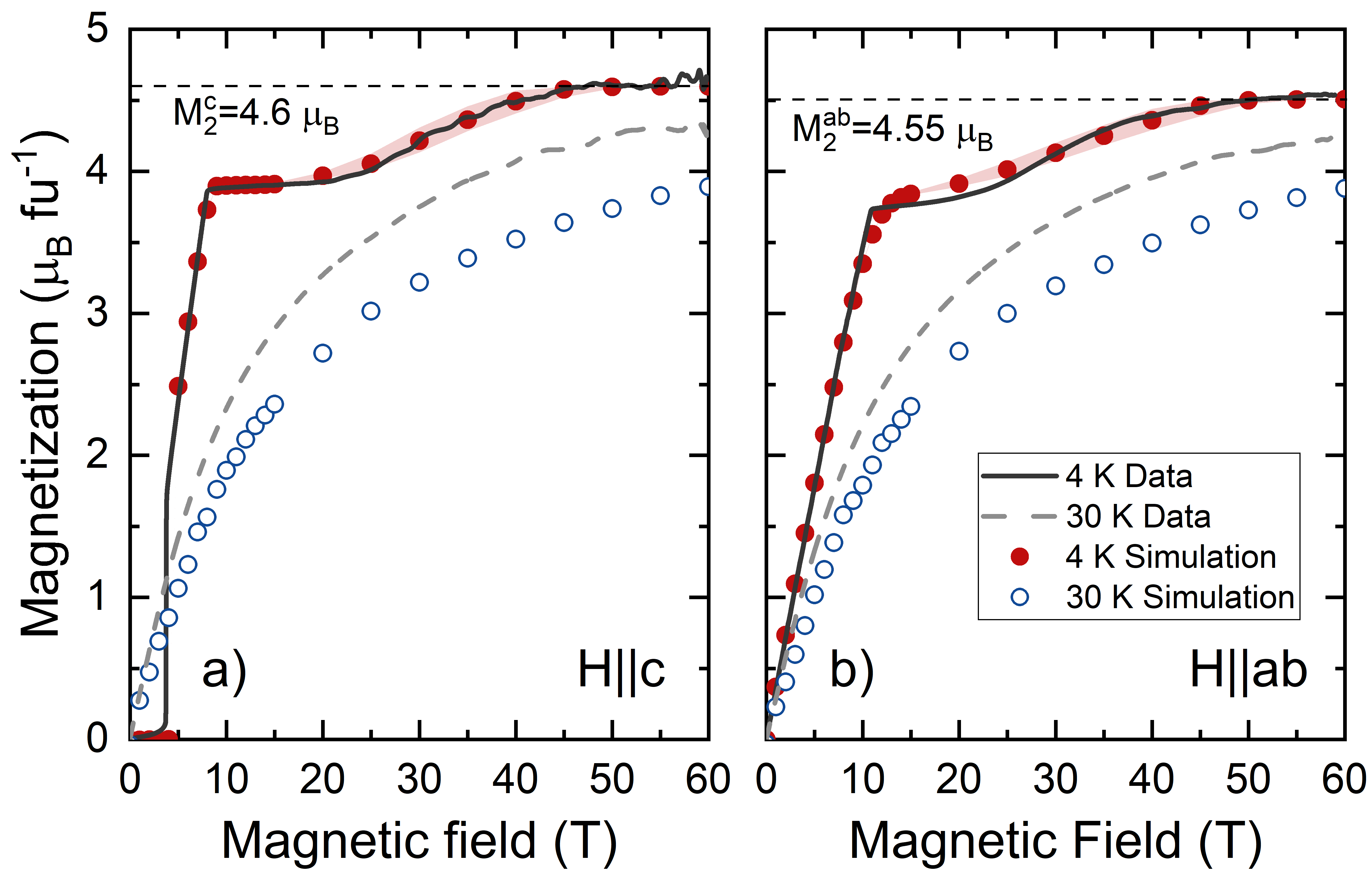}
\caption{High-field magnetization data of MnBi$_{2}$Te$_{4}$ at $T=4$~K (black solid line) and $T=30$~K (gray dashed line) compared to results from classical Monte Carlo simulations at $T=0.1$~K (red solid circles)  and $T=30$~K (blue empty circles). Panel a) has the field applied along the $c$-axis and panel b) in the $ab$-plane. Heisenberg model simulation parameters are described in the text with $x=0.038$ and $SJ'=$ 1.2~meV. The shaded red area corresponds to the range of magnetization in the MC simulations when $SJ'$ is varied by $\pm$ 10 \%.}
\label{MC}
\end{figure}

Fig.~\ref{MC}a) shows that reasonable agreement between the simulations and the $H \parallel c$ data is achieved for $x=0.038$ and $SJ'= 1.2$~meV. We note that the intrablock AF coupling strength $J'$ is found to be much larger than other magnetic couplings in the system. As detailed by density-functional theory calculations described below, a major contributor to $J'$ is AF superexchange coupling mediated through the linear next-nearest neighbor (NNN) Mn$_{\rm Mn}$--Te--Mn$_{\rm Bi}$ bond.  The Mn$_{\rm Bi}$ ions, each coupled to at most three NNN Mn$_{\rm Mn}$ ions, reside in a molecular field with strength $H_{\rm MF}$ $\approx 3SJ'/(g\mu_{\rm B}) \approx$ 34~T which sets the characteristic field scale for full saturation.  

Leaving all Heisenberg parameters fixed, Fig.~\ref{MC}b) compares magnetization simulations to the data when the field is applied the $ab$-plane. While the simulations provide a reasonable agreement for this field-direction, there are some subtle differences that could arise from weak anisotropies, quantum effects or longer-range interactions that are not accounted for in the model/simulations.

\begin{figure}
\includegraphics[width=3. in]{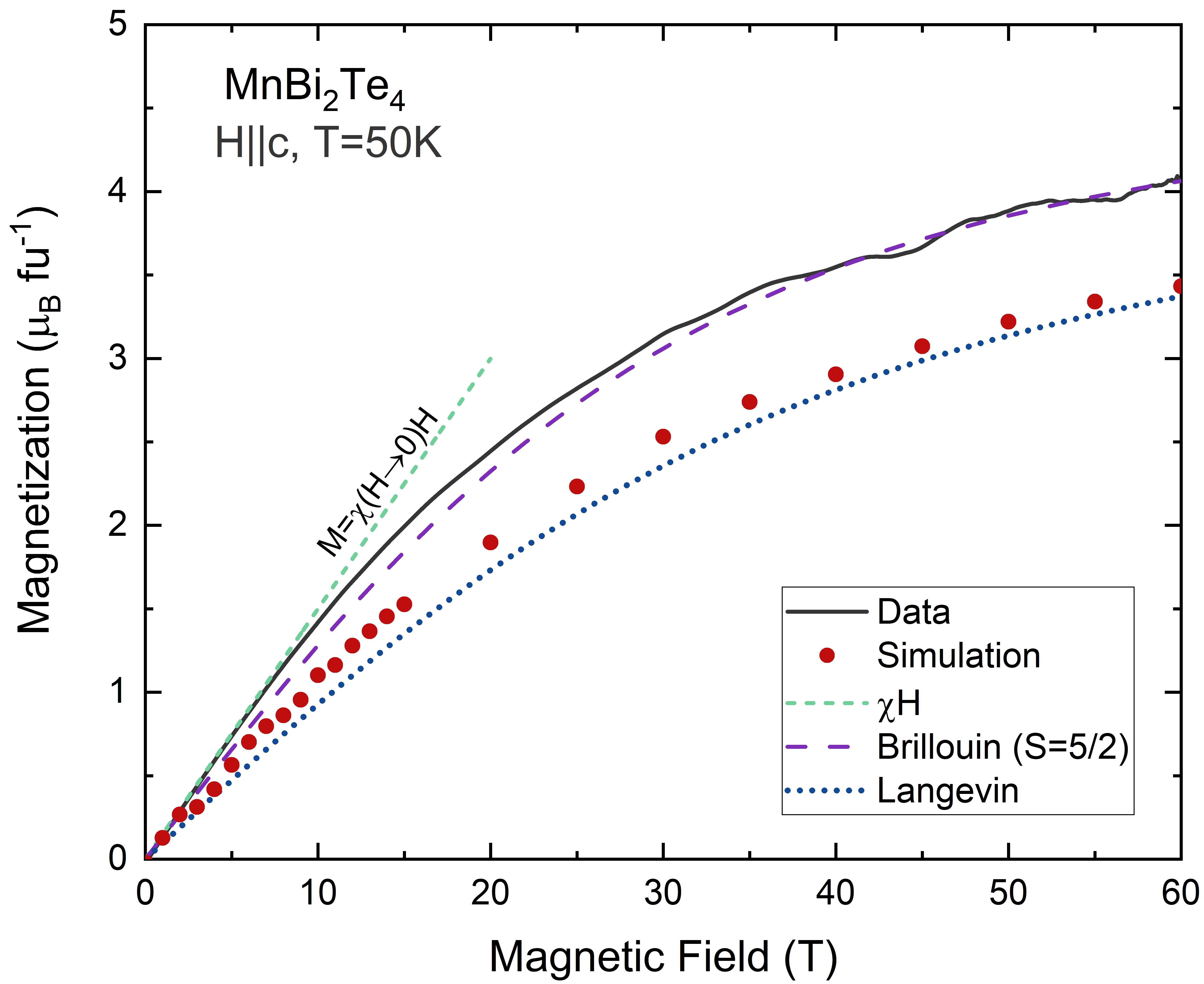}
\caption{High-field magnetization data of MnBi$_{2}$Te$_{4}$ (black line) measured at 50~K compared to CMC simulations (red circles).  For comparison, the quantum mechanical $S=5/2$ Brillouin function (purple dashed line) and classical Langevin function (blue dotted line) are shown.  Finally, the light green dashed line corresponds to the linear magnetization extrapolated from low-field susceptibility data \cite{Yan19}, $M=\chi H$, where $\chi=$ 0.15 $\mu_{\rm B}$ T$^{-1}$ fu$^{-1}$ at 50~K.}
\label{50K}
\end{figure}

Figs.~\ref{MC}a) and b) also show a comparison of CMC simulations and data at $T=30$~K. Here, the CMC simulations underestimate the measured magnetization. This discrepancy with the classical spin dynamics model is expected at higher temperatures where spin fluctuations become more important. For example, in the paramagnetic limit at temperatures well above $T_{\rm N}$, the magnetization data should follow the quantum mechanical $S = 5/2$ Brillouin function, whereas the classical simulations follow the Langevin function. Fig.~\ref{50K} compares the $H \parallel c$ data at 50 K with MC simulations, as well as the $S = 5/2$ Brillouin and Langevin functions calculated with a magnetic moment of 4.6~$\mu_{\rm B}$.  Examination of this plot suggests that MBT is not in the paramagnetic limit even at $T = 50~{\rm K} \approx 2T_{\rm N}$. The high temperature simulations also predict isotropic magnetization for $T > T_{\rm N}$ (see Fig.~2) which is expected because the temperature scale of the single-ion anisotropy is small $SD/k_{\rm B} \approx$ 1~K. Experimentally, we find a few percent difference in the absolute magnetization scale for fields applied along $ab$ and $c$ (see Fig.~\ref{data}) which may arise from orbital hybridization effects that are not considered in our model.

\section{Analysis of magnetization data for MST}

\begin{figure}
\includegraphics[width=3.3 in]{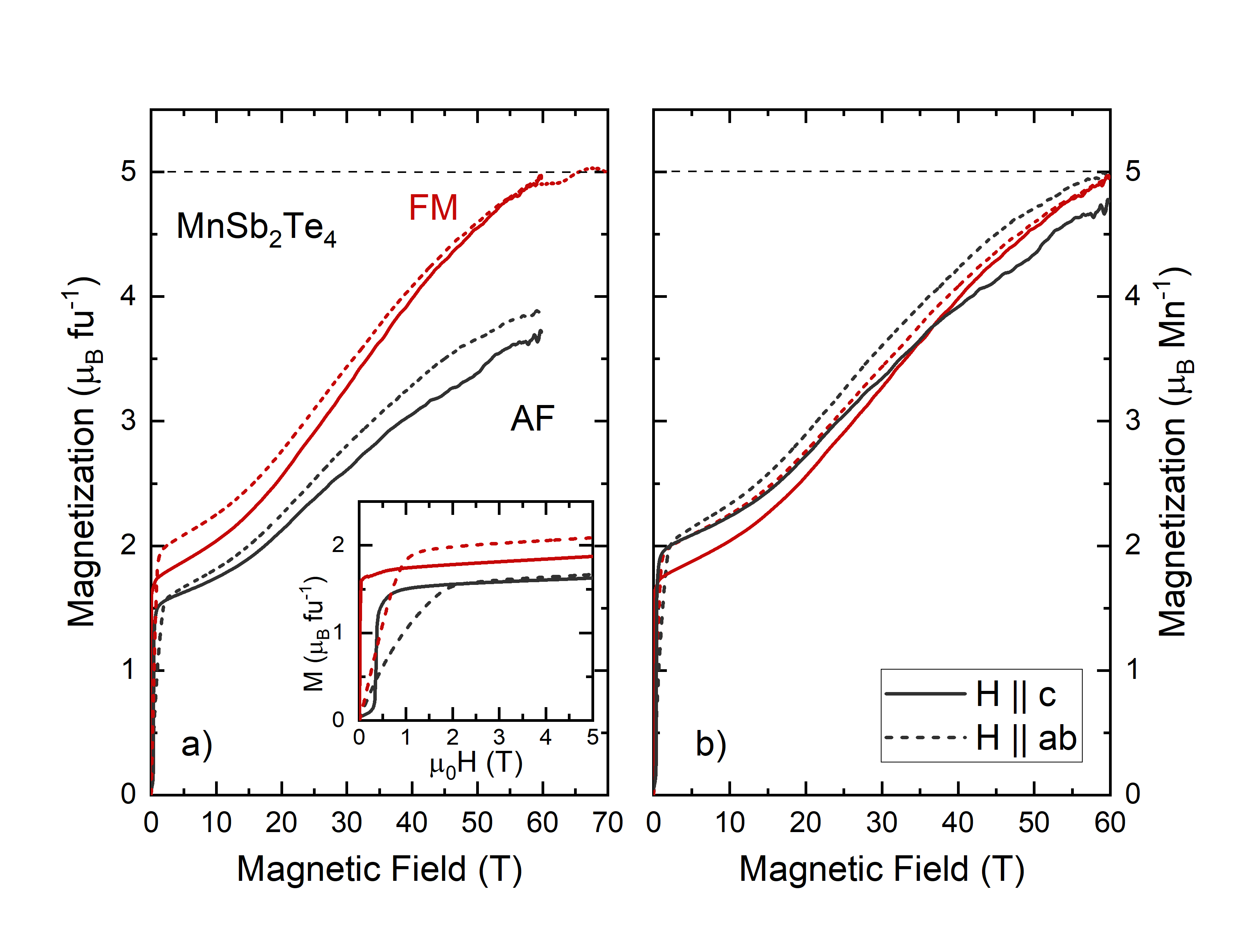}
\caption{ High-field magnetization data of nominal MnSb$_{2}$Te$_{4}$ samples that adopt both MST-FM (red lines) and MST-AF (black lines) ground states with chemical formulas of Mn$_{1.00(4)}$Sb$_{2.09(2)}$Te$_{3.91(2)}$ and Mn$_{0.778(5)}$Sb$_{2.222(6)}$Te$_{4.00(1)}$, respectively.  Magnetization data are normalized a) per formula unit and b) per Mn. Measurements are performed at 4~K and with the applied field $H \parallel c$ (solid lines) and $H \parallel ab$ (dashed lines). The inset to panel a) zooms in on the low-field region.}
\label{MST_data}
\end{figure}

For MST, previous experimental evidence supports much higher levels of antisite defects.~\cite{Murakami19, Liu20, zhou20} In particular, the strongly reduced low-field "saturation" magnetization of $< 2~\mu_{\rm B}$ in MST suggests a significant concentration of AF coupled antisite defects are present that act to compensate the net magnetic moment~\cite{Yan19_MST, Liu20}. Furthermore, depending on the details of the sample preparation MST may adopt either a FM or AF magnetic ground state~\cite{Liu20}, and it is suspected that different defect configurations control which magnetic ground state is adopted. Here, the magnetization of both MST-FM and MST-AF samples are measured up to fields of 60~T. Fig.~\ref{MST_data}a) reveals that both samples quickly saturate to a low-field magnetization plateau near 2~$\mu_{\rm B}$ followed by gradual recovery of significant missing magnetization at higher fields. This observation is consistent with the current picture of ferrimagnetic septuple blocks that become fully polarized in large fields. Fig.~\ref{MST_data}a) also indicates that the magnetization per formula unit of the MST-AF sample is generally lower, as we show below, which can be traced back to differences defect configuration between MST-FM and MST-AF samples.

We now estimate the defect configuration of the two MST samples from EDS and magnetization data. EDS measurements reveal that the MST-FM sample is nearly stoichiometric in Mn (Mn$_{1.00(4)}$Sb$_{2.09(2)}$Te$_{3.91(2)}$, $y=0.0$) whereas the MST-AF sample is Mn deficient (Mn$_{0.778(5)}$Sb$_{2.222(6)}$Te$_{4.00(1)}$, $y=0.22$).  The Mn deficiency explains the suppression of the magnetization per formula unit of the MST-AF sample.  Using the same approach as for MBT, the antisite mixing concentration $x$ is estimated by first considering the bare Mn moment size. Fig.~\ref{MST_data}b) plots the same data as Fig.~\ref{MST_data}a), but with units of $\mu_{\rm B}$ per Mn rather than per fu. This rescaling gives some confidence in our EDS determined stoichiometries since the moment per Mn are very similar in MST-FM and MST-AF samples. However, this rescaling suggests that the magnetization of both MST samples are close to reaching saturation by 60$-$70~T and $m_0 > 5~\mu_{\rm B}$. This is a somewhat surprising result, since DFT predicts a similar moment of ~4.7~$\mu_{\rm B}$ for MST and MBT. One possibility is that magnetic impurity phases such as MnSb may exist in MST samples, where a few kOe is needed to saturate the MnSb moment,~\cite{MnSb_impurity} which might lead us to over estimate the magnetization of MST.
Estimates of $M_1$ and $M_2$ for both MST-AF and MST-FM are shown in Table \ref{table3}. Using the EDS data and Eqns.~3 and 4 allows us to estimate that $x=0.11$ for the MST-AF and 0.15 for the MST-FM, the former is consistent with the neutron scattering study performed on single crystals from the same batch.~\cite{Liu20}

To develop a Heisenberg model for the MST samples, we again estimate the exchange parameters based on the Stoner-Wohlfarth model.~\cite{swmodel} However, unlike MBT, the spin-flop transition is absent and instead a spin-\textit{flip} transition is found at $H_1^c$ for MST-AF. This occurs in the regime of $D/zJ_c > 1$, where the spin-flip field is $g \mu_{\rm B}H_{flip} = g \mu_{\rm B}H_1^c = zSJ_c$ giving
\begin{equation}
SJ_c = g\mu_{\rm B}H_1^c/z.
\label{eqnJc_MST}
\end{equation}
The relation $g \mu_{\rm B}H_1^{ab}=2zSJ_c+2SD$ provides an estimate of $SD$ in this regime
\begin{equation}
SD = g\mu_{\rm B}(H_1^{ab}-2H_1^c)/2.
\label{eqnD_MST}
\end{equation}
For MST-AF, the inset in Fig.~\ref{MST_data}a) allows for an evaluation of $H_1^c \approx 0.5$ T and  $H_1^{ab} \approx 2.2$~T from which we obtain $SJ_c \approx 0.01$ meV and  $SD \approx 0.07$ meV. For MST-FM, $H_1^c$ (which is essentially zero) is determined primarily by the energetics of domain boundaries, which are difficult to estimate. Given the rather small value of $J_c$ for the MST-AF and the proximity to FM order, it is not unreasonable to assume that $J_c \approx 0$ for MST-FM. In this approximation we obtain $SD \approx g \mu_{\rm B}H_1^{ab}/2 \approx  0.07$~meV.  We note that the values $SD$ obtained for MST-AF, MST-FM and MBT are nearly the same, in agreement with DFT calculations.~\cite{MBT_DFT}

\begin{figure}
\includegraphics[width=3.1 in]{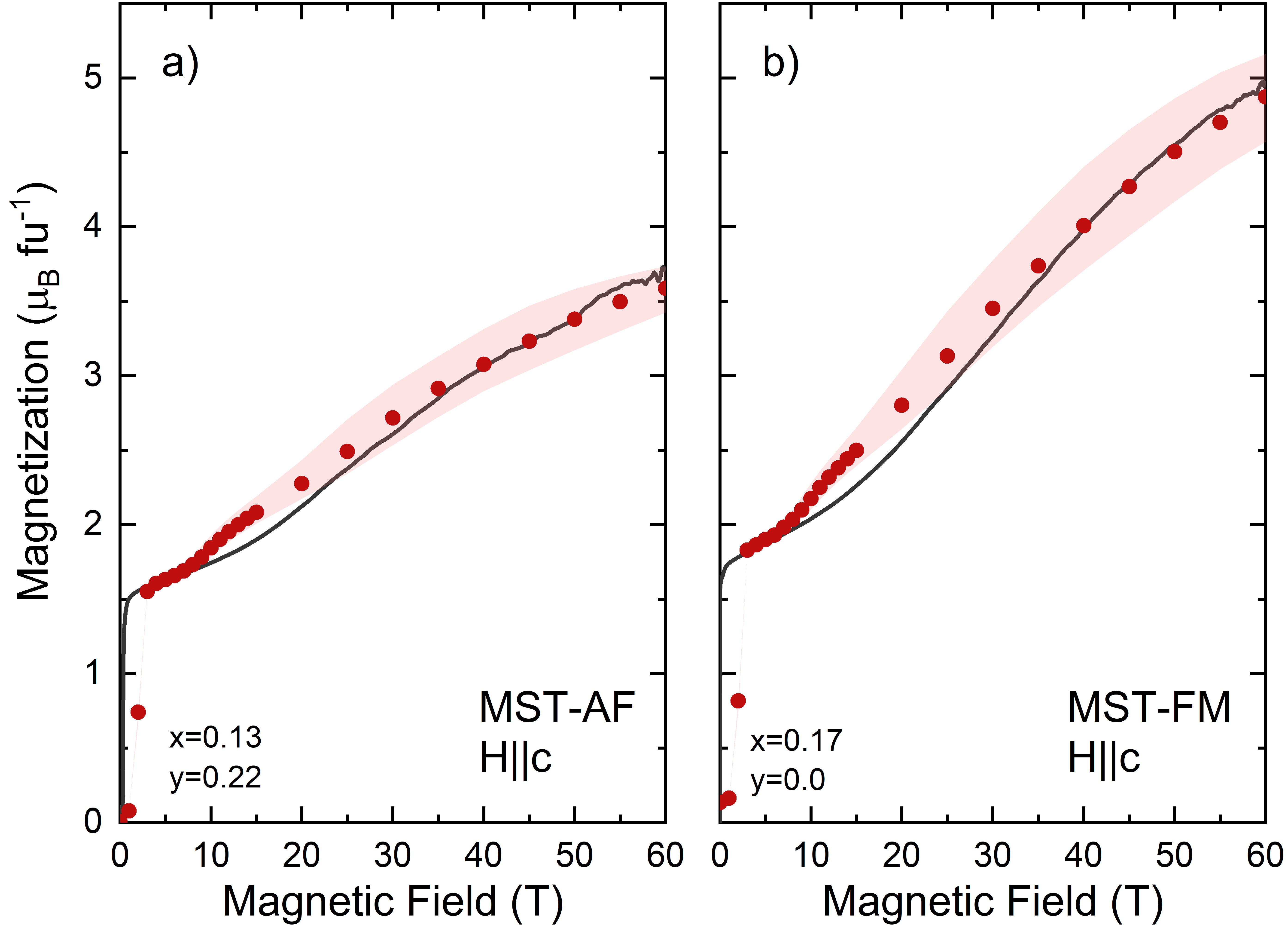}
\caption{a) High-field magnetization data of a) MST-AF~(Mn$_{0.778(5)}$Sb$_{2.222(6)}$Te$_{4.00(1)}$) and b) MST-FM~(Mn$_{1.00(4)}$Sb$_{2.09(2)}$Te$_{3.91(2)}$) samples at $T=4$ K (black solid line) compared to results from classical Monte Carlo simulations at $T = 0.1$~K (red solid circles). The field applied along the $c$-axis. Heisenberg model simulation parameters are described in the text with $SJ' =$ 2.1 meV and $x=0.13$, $y=0.22$ (AF) and  $x=0.17$, $y=0$ (FM).  The shaded red area corresponds varying $SJ'$ by $\pm 15$\%.}
\label{MC_MST}
\end{figure}

We use the approximate defect configuration and Heisenberg parameters as a starting point for our CMC simulations for both MST samples.  Based on uncertainty of saturation magnetization, we fix $y$ to its EDS value.  For MST-AF, Fig.~\ref{MC_MST}a) shows that reasonable agreement is obtained for $x=0.13$, $m_0=$ 4.9 $\mu_{\rm B}$ and $SJ' = 2.1$ meV. For MST-FM, Fig.~\ref{MC_MST}b) CMC simulations are shown for $x=0.17$, $m_0=$ 5.3 $\mu_{\rm B}$ and $SJ' = 2.1$~meV.  However, the CMC simulations for MST samples generally provide less satisfactory agreement with the data as compared to MBT. The overshooting of the simulated magnetization between 10 -- 30~T suggests that additional AF interactions are present (for example, between Mn$_{\rm Sb}$ ions either within the septuple block or across the van der Waals gap). The complexity of the intrablock magnetic interactions and the overall larger value of $J'$ for MST as compared to MBT is supported by DFT calculations described in the next section.

\section*{\emph{Ab initio} calculation of intrablock exchange coupling in MBT and MST}

\begin{figure}[ht]
\centering
\begin{tabular}{c}
\includegraphics[width=.80\linewidth,clip]{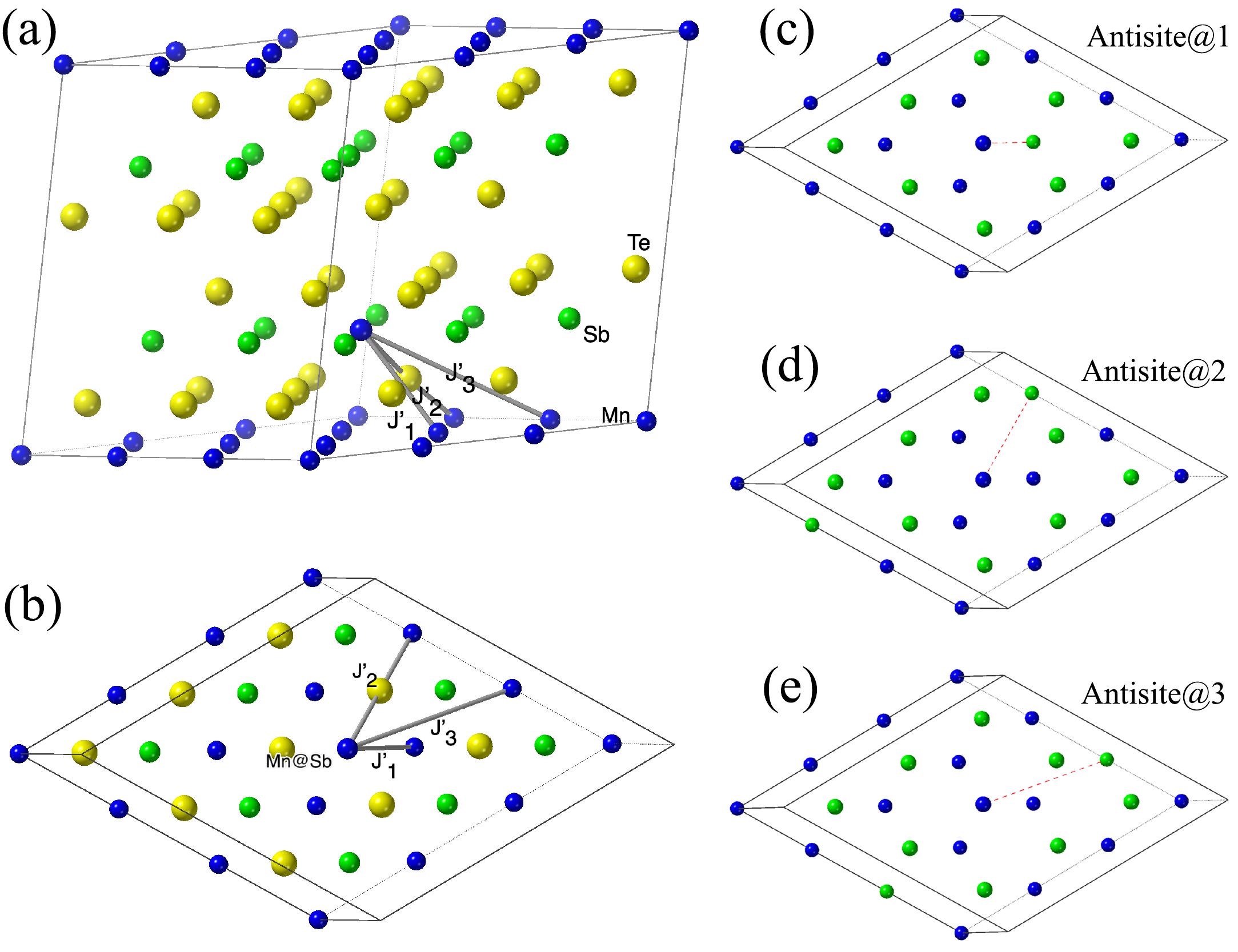}%
\end{tabular}%
\caption{(a) Schematic representation of the $3\times3\times1$ supercell used in the DFT calculation for  MST.  One of the 18 Sb atoms is substituted by Mn. Mn, Sb, and Te atoms are indicated with blue, brown, and yellow spheres, respectively. The first three nearest intrablock Mn$_\text{Sb}$-Mn$_\text{Mn}$ exchange parameters are labeled as $J'_1$, $J'_2$, and $J'_3$, respectively, , are also shown from the perspective along the c-axis in (b).  (c)-(e) show the three different scenarios for the formation of antisite-defect configurations at nearest-, next-nearest, and next-next-nearest neighbor positions, respectively (Te atoms are not shown). 
 }
\label{fig:xtal}
\end{figure}

\vspace{-10pt}

\begin{table}[hbt]
  \caption{Calculated formation energies $E_\text{FN}$ and intrablock magnetic energies $E_\text{M}$ of antisite defects in MBT and MST. The $E_\text{M}$ values of a Mn$_\text{Sb/Bi}$ defect are also listed to compare. Both $E_\text{FN}$  and $E_\text{M}$  are in the unit of meV/Mn$_\text{Bi/Sb}$. Calculations were carried out in DFT+$U$ using $U = 4$~eV. $E_\text{M}$ is calculated as the energy difference between the AFM and FM intrablock configurations, $E_\text{AF}-E_\text{FM}$. Positive (negative) $E_\text{M}$ values correspond to FM (AF) intrablock couplings.}
\label{tbl:mi}
\bgroup
\def\arraystretch{1.1}
\begin{tabular*}{\linewidth}{l @{\extracolsep{\fill}} crrrrr}
\hline\hline
\multirow{2}{*}{Configuration}           &  &   \multicolumn{2}{c}{MST}  &    &   \multicolumn{2}{c}{MBT} \\  \cline{3-4} \cline{6-7}
           & & $E_\text{FN}$   & $E_\text{M}$    &   & $E_\text{FN}$   & $E_\text{M}$  \\ \hline 
Mn$_\text{Sb/Bi}$ & &     &    11.6 & &          &  21.11 \\
Antisite@infinity  & & 689 &    & &     1254  &   \\
Antisite@1        & & 115 &   -8.21 & &     363  & -6.74  \\
Antisite@2        & & 115 &   -0.03 & &     315  & -2.50 \\
Antisite@3        & & 135 &   -2.07 & &     375  & -1.15 \\
\hline\hline
\end{tabular*}
\egroup
\end{table}

DFT+$U$~(where $U$ is the additional onsite coulomb interactions) calculations were carried out to investigate the defect-driven intrablock exchange couplings in MST and MBT. To simulate the additional Mn $d$-orbital on-site electron-electron repulsion not accounted for in DFT calculations, we apply $U$ = 4 -- 5~eV, which was found to better describe the magnetic interactions in these systems~\cite{Li20}.
To simulate defects, we use a supercell that corresponds to a $3\times3\times1$ superstructure of the original primitive (FM) cell, as shown in Fig.~\ref{fig:xtal}.
We first investigate the coupling between a single Mn$_\text{Sb/Bi}$ defect and Mn$_{\rm Mn}$ neighbors in the adjacent Mn layer. Then we further substituted one Sb/Bi atom on various Mn$_\text{Mn}$ sites on the adjacent Mn layer to study the effects of additional Sb$_\text{Mn}$/Bi$_\text{Mn}$ antisite vacancies on the intrablock coupling.

We estimate the intrablock couplings by employing both a linear response method as implemented in the LMTO-ASA-GF code~\cite{ke2013prb,ke2017prb}, and total energy calculations using the Vienna \textit{ab initio} simulation package (\textsc{vasp})~\cite{kresse1993prb,kresse1996prb}. Calculations are performed within the generalized gradient approximation using the exchange-correlation functional of Perdew, Burke, and Ernzerhof.~\cite{pbe} In \textsc{vasp}, the nuclei and core electrons are described by the projector augmented-wave potential~\cite{kresse1999prb}, and the wave functions of valence electrons are expanded in a plane-wave basis set with a cutoff energy of up to \SI{350}{eV}. Table~\ref{tbl:mi} summarizes the formation energies and the intrablock magnetic energies of the corresponding configurations calculated using \textsc{vasp}.

First, we start with MST, in which the defects are more dominant than in MBT, and substitute one of 18 Sb atoms with Mn in the supercell. The first three nearest exchange couplings between this Mn$_\text{Sb}$ magnetic defect and the neighboring Mn$_\text{Mn}$ layer are denoted as $J'_1$, $J'_2$, and $J'_3$, respectively, in Fig.~\ref{fig:xtal}(a) and (b). We calculated the intrablock exchange couplings at $U = 4$~eV.  

ASA-GF calculations find that the NN $J'_1$ interaction between Mn ions separated by $\sim$ 4.5 \AA~and a Mn--Te--Mn angle close to 90$^\circ$ gives a net FM contribution to the intrablock coupling. The NNN $J'_2$ interaction at $\sim$6.2 \AA~corresponds to AF superexchange with an Mn--Te--Mn angle close to $180\degree$. The third nearest-neighbor $J'_3$ has a slightly AF coupling.  Thus, the net intrablock coupling will depend on the relative strengths of the FM and AF couplings. To better quantify the strength of the net intrablock coupling, we fully relax the atomic coordinates while keeping the lattice parameters (adopted from experiments) fixed and calculate the total energies of FM and AF intrablock configurations using \textsc{vasp}. In contradiction with experimental results, the net coupling of single Mn$_{\rm Sb}$ defect with FM coupling to the FM layer is found to have lower energy than the AF coupling by 11.6~meV/Mn$_\text{Sb}$.

A possible way to reconcile this contradiction is to consider the possibility that antisite defects are correlated or bound in real space. To account for this possibility, we calculate the intrablock coupling associated with a coupled pair of antisite defects and find that Sb$_\text{Mn}$ defects promote the AF intrablock coupling. Within the given supercell, three possible scenarios were explored, where Sb$_\text{Mn}$-Mn$_\text{Sb}$ antisite pairs are placed at NN, NNN, and 3$^{rd}$ NN positions, as shown in Fig.~\ref{fig:xtal}(c--e). These defect configurations are hereafter referred to as antisite@{$i=$1, 2, 3}. A comparison of the calculated formation energies of antisite@$i$ configurations to the single Mn$_\text{Sb}$ and Sb$_\text{Mn}$ defects (antisite@infinity) show that bound antisite pairs are energetically favored. Furthermore, the antisite@1 and antisite@2 have similar lower energy than antisite@3.

As shown in Table~\ref{tbl:mi}, all three antisite configurations favor AF structure, demonstrating that vacancies in the Mn layer promote AF intrablock coupling.  Interestingly, among them, antisite@1 has the strongest AF coupling, while antisite@2 is the least. To compare to the Heisenberg model and compare to CMC simulations, we estimate that $SJ'$ $\approx$ $E_\text{M}$/$M$ = 1.8~meV for MST, where $M$ = 4.5~$\mu_{\rm B}$ is the moment obtained from DFT.

In a naive picture, a removal of one out of three of FM $J'_1$ bonds in antisite@1 will favor a net AF coupling when compared to the original single Mn$_\text{Sb}$ defect case. However, in general, besides eliminating a corresponding exchange bond, Sb$_\text{Mn}$ may also affect the coupling of other pairs. For example, removal of 1/3 of the original AFM $J'_2$ bonds in antisite@2 should promote stronger FM coupling. However, Table \ref{tbl:mi} shows that the overall coupling for antisite@2 is AF as well. From our limited exploration of defect-mediated coupling, we can only conclude that correlated antisite defects favor a net AF coupling in all configurations studied.

For MBT, an Mn$_\text{Bi}$ defect shows an even stronger FM intrablock coupling with the Mn layer than Mn$_\text{Sb}$ in MST. Similarly, correlated antisite defect pairs stabilize the AF intrablock coupling in a similar fashion as MST. The formation energies of antisite pairs are much larger in MBT than in MST, consistent with the experimental fact that antisite defects are more prevalent in MST than in MBT. In comparison to MST, the antisite@2 configuration, which is less AF-like than antisite@1, is more favorable than antisite@1 in MBT. Overall, Table \ref{tbl:mi} is consistent with the experimental result that the AF intrablock coupling is generally smaller in MBT.

From DFT, we conclude that (1) the magnetic vacancies on the Mn layer, Sb$_\text{Mn}$ or Bi$_\text{Mn}$, promote  AF intrablock coupling, and (2) MST has a stronger intrablock AF coupling than MBT, consistent with experiments.  However, one caveat in these DFT results is that the absolute and relative values of these couplings are highly dependent on the value of $U$ and the choice of functional. On the one hand, for both MST and MBT, the coupling of a single defect to the Mn layer becomes more FM with a larger $U$ parameter, sharing a similar trend as DFT studies of the intralayer couplings~\cite{Li20}. At $U = 5$~eV, all three antisite cases considered above become FM in both MBT and MST unless more antisite defects and Bi$_\text{Mn}$/Sb$_\text{Mn}$ vacancies are introduced to stabilize an AF intrablock configuration. On the other hand, different functionals can change the relative strengths of pairwise FM and AF interactions. The defect distributions and configurations in real systems are likely more complicated than the limited scenarios we considered here and deserve a more comprehensive future study, ideally, in a beyond-DFT and parameter-free fashion~\cite{ke2021ncm}.

\section{Conclusion}
In summary, we reveal the hidden magnetization in MBT and MST materials by applying high magnetic fields. This shows, to varying degrees, that MXT materials have partially compensated magnetization due to strong AF coupling of magnetic antisite defects to the main Mn layer. This confirms a picture of MXT samples where each septuple block adopts an overall ferrimagnetic configuration consistng of both a uniform and staggered magnetization component, that might play a role in determinations of quantum anomalous Hall state and surface state properties in MXT.

DFT studies also highlight that competing FM and AF intrablock interactions between Mn$_\text{X}$ and Mn$_\text{Mn}$ moments are present and the net AF interaction is only obtained after considering the configuration of various defect states. Overall, intrablock ferrimagnetism is favored in the case where antisite pairs are bound together in close proximity. We investigated the role of random and correlated magnetic defects in the magnetism of MXT materials, an extension of this study might explain how different chemical and magnetic configurations in MST-FM and MST-AF determine the overall sign of the interblock coupling.

Also, one can connect our results of the magnetic interactions of antisite defects in MXT to the case of dilute Mn substitutions in the Bi$_2$Te$_3$ topological insulator.~\cite{Bi2Te3_Mn} For the dilute magnetic TI, global ferromagnetism is promoted which clearly indicates that the net intralayer, intrablock, and interblock couplings are all FM. With respect to the intrablock coupling, it might seem inconsistent that FM coupling is found between bilayers in dilute magnetic TI, since the local bonding is quite similar to MXT compounds. However, our results show that the strength and sign of the net coupling is highly dependent on the local configuration of magnetic defects, which is vastly different in the two cases. Ultimately, these results outline a clear need for first-principles calculations where the magnetic interactions are studied under various assumptions about the configuration of magnetic defects and vacancies, in order to understand the role of complex magnetic configuration on the band topology of magnetic metals.

\section{Acknowledgements}
We thank Dr.~Johanna Palmstrom, Dr.~Laurel Stritzinger and Dr.~John Singleton for their assistance with the measurement in the Pulsed Field Facility at LANL, and we appreciate the comments and discussions that Dr.~John Singleton provide that greatly improved the manuscript. This work was supported by the Center for the Advancement of Topological Semimetals, an Energy Frontier Research Center funded by the Department of Energy, Basic Energy Sciences, under Contract No. DE-AC02-06CH11357. L.K. acknowledges the support from the U.S. DOE Early Career Research Program. A portion of this work was performed at the National High Magnetic Field Laboratory, which is supported by the National Science Foundation Cooperative Agreement No.~DMR-1644779 and the state of Florida. Research performed at ORNL is supported by the U. S. Department of Energy, Office of Science, Basic Energy Sciences, Materials Sciences and Engineering Division.

\end{document}